\documentclass[final,5p,times,twocolumn]{elsarticle}

\usepackage{hyperref}

\journal{Computational Materials Science}

\begin{document}

\begin{frontmatter}

\title{Unsupervised learning of atomic environments from simple features}

\author[mymainaddress,mysecondaryaddress]{Wesley F. Reinhart\corref{mycorrespondingauthor}}

\cortext[mycorrespondingauthor]{Corresponding author}
\ead{reinhart@psu.edu}

\address[mymainaddress]{Department of Materials Science and Engineering, Pennsylvania State University, University Park, PA, 16802, USA}
\address[mysecondaryaddress]{Institute for Computational and Data Sciences, Pennsylvania State University, University Park, PA, 16802, USA}

\begin{abstract}

I present a strategy for unsupervised manifold learning on local atomic environments in molecular simulations based on simple rotation- and permutation-invariant three-body features.
These features are highly descriptive, generalize to multiple chemical species, and are human-interpretable.
The low-dimensional embeddings of each atomic environment can be used to understand and quantify messy crystal structures such as those near interfaces and defects or well-ordered crystal lattices such as in bulk materials without modification.
The same method can also yield collective variables describing collections of particles such as for an entire simulation domain.
I demonstrate the method on colloidal crystallization, ice crystals, and binary mesophases to illustrate its broad applicability.
In each case, the learned latent space yields insights into the details of the observed microstructures.
For ices and mesophases, supervised classifiers are trained based on the learned manifolds and directly compared against a recent neural-network-based approach.
Notably, while this method provides comparable classification performance, it can also be deployed on even a handful of observed environments without labels or \textit{a priori} knowledge.
Thus, the current approach provides an incredibly versatile strategy to characterize and classify local atomic environments, and may unlock insights in a wide variety of molecular simulation contexts.
\end{abstract}

%\begin{keyword}
%\end{keyword}

\end{frontmatter}

\section{Introduction}\label{sec:intro}

% TODO: polish this intro paragraph
Functional materials whose microscopic structure influences their macroscopic properties continue to become more available and more complex as a result of new and improved synthesis technologies.
A wide variety of molecular simulation and, lately, machine learning techniques have been applied to assist in navigating this expansive design space.
In order to apply rational and data-driven design schemes, quantitative descriptions of the microstructure are required.
Many different methods have been used to characterize local atomic environments, but no single approach works for all systems or all purposes.
For instance, in adaptive sampling schemes the collective variables need to be computed every timestep and therefore must be inexpensive, while a larger computational budget is viable for strictly postprocessing simulation data.
% TODO: decide how to deal with this trailing thought

For order-disorder phase transitions, an \textit{order parameter} is a scalar value that quantifies the position in phase space based on the real-space coordinates of particles in the system.
Defining a mathematical function which reduces the dimensionality from $n$ to 1 provides major advantages for both qualitative and quantitative understanding of a system \cite{sethna2021statistical}.
A simple example is the nematic order parameter for liquid crystals \cite{priestly2012introduction}.
For polymorphic crystal-forming systems, more sophisticated, higher-dimensional order parameters are needed to capture the essential features, though there have been some recent efforts focused on unsupervised quantification of global order in such cases.\cite{Jadrich2018,Martiniani2019}.
Perhaps the most common crystalline order parameters are the bond order parameter \cite{Steinhardt1983,Lechner2008}, which have provided insight into molecular simulations for an extraordinary range of physical systems over the last several decades.
The bond order parameters are a collection of specific combinations of the spherical harmonic basis functions which are invariant to rotations of the atomic coordinates.
Taking a sum over nearest neighbors also makes them invariant to permutations of the order of atoms.
These two properties, rotation and permutation invariance, are essential to any description of local atomic environment \cite{Bartok2013}.

In the past, I \cite{Reinhart2017Machine,Reinhart2017Multi,Reinhart2018Automated} and others \cite{Honeycutt1987,blatov2000search,Larsen2016,Xie2018} have explored graph-based descriptors, because they can discriminate between crystal polymorphs more accurately than continuous order parameters.
However, the graph-based descriptors have two major drawbacks: (i) they struggle with liquids (or other disordered states) since the topology of a liquid is essentially random, and (ii) they are not typically differentiable -- or when they are, as in recent work with graph convolutional networks such as Ref.~\cite{Xie2018}, they instead require a large volume of training data to learn suitable weights.

While the above methods have mostly been applied for unsupervised classification schemes, there has also been a substantial amount of work published around \textit{atomic fingerprints} for regressions with supervised learning \cite{Behler2011,Rupp2012,Bartok2013,Li2015,Botu2015,Huan2017}.
A collection of similar representations has been developed specifically for use in conjunction with neural networks to make fast and accurate predictions of many-body forces and energies after being trained on more expensive \textit{ab initio} calculations \cite{Xie2018,Behler2007,Artrith2017,Zhang2018Deep,Zhang2018End,Chandrasekaran2019,Mailoa2019}.
These have been fairly successful in representing atomic environments for molecular simulation, but typically require tens of thousands of labeled training samples to produce accurate results.
This is not insurmountable for regression of atomic energies, since obtaining labeled data is as simple as computing the energies using standard simulation software (though this may carry a high computational cost).
However, labeled structure classification data is difficult or impossible to obtain since there is no ground truth for thermalized atomic environments.

Here I develop a descriptor for the local atomic environment based on a simple set of rotation-invariant features plus a permutation-invariant pooling transformation which captures fine details (e.g., distinguishes between crystal polymorphs) but is also differentiable.
For the rotation-invariant features, I choose simple three-body terms: interparticle separation distance, bond length, and bond angle.
To achieve permutation invariance, I create Gaussian expansions (i.e., fuzzy histograms) of these features over each local environment.
While this expansion makes the proposed method slower than other, simpler strategies like Common Neighbor Analysis\cite{Honeycutt1987} or Polyhedral Template Matching,\cite{Larsen2016} which only need to evaluate a few hand-crafted features, it offers additional opportunities for feature learning through unsupervised manifold learning.
This unsupervised approach is can provide insight into complex microstructure with zero training data, or provide robust classification for large systems when labeled training data is available.
It is sensitive to minute differences between crystal polymorphs but can also handle distinction between liquid and crystalline environments, or even different amorphous environments.

I demonstrate the results of the proposed method on three different test systems: colloidal crystals, ices, and binary mesophases.
In the first example, the method is used to characterize small colloidal crystal domains (taken from Ref.~\cite{Howard2018Evaporation} and previously studied in Refs.~\cite{Reinhart2017Machine,Reinhart2018Automated}) using unsupervised manifold learning.
The discovered latent space is found to exhibit physical meaning related to coordination number, surface character, and crystal symmetry.
Collective variables are automatically obtained which can be closely matched to intuitive, hand-selected order parameters in the system.
In the second example, the method is used to classify ice structures taken from Ref.~\cite{Defever2019}, with the results comparing favorably to classification by a deep neural network which was trained on $10 - 100 \times$ as many samples.
Furthermore, the distinction between bulk and surface phases which required special accommodations in the neural network training does not confuse my new scheme.
Finally, the method is demonstrated on binary mesophase structures (also taken from Ref.~\cite{Defever2019}) and again compares favorably to prior results from supervised learning while requiring less training data and providing additional insight through the provision of a structural latent space.

\section{Methods} \label{sec:methods}

\begin{figure}
\center
\includegraphics[width=9cm]{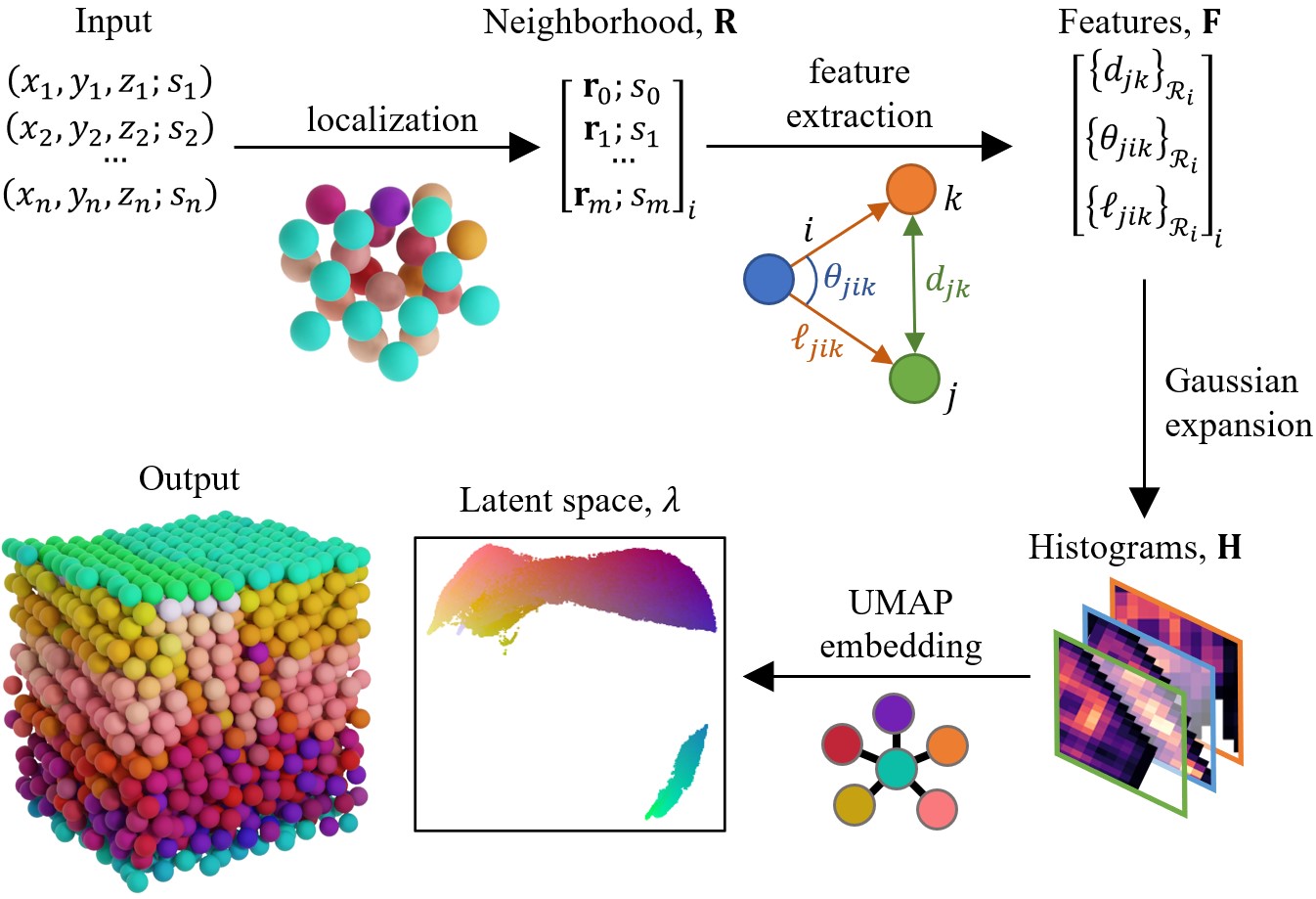}
\caption{
Schematic of the proposed unsupervised learning scheme using rotation-invariant features $\mathbf{F}$, a Gaussian expansion to obtain the permutation-invariant $\mathbf{H}$, and unsupervised manifold learning to obtain a low-dimensional latent space $\lambda$.
Include sample neighborhood and histograms.
Renderings performed using Blender.
\label{fig:schematic}
}
\end{figure}

\subsection{Rotation and permutation invariance}\label{sec:features}

Here I develop a scheme to obtain simple, interpretable, rotation-invariant features.
For each particle $i$, a local neighborhood matrix $\mathbf{R}_i$ is defined according to an isotropic cutoff radius, such that the rows $\mathbf{R}_{i,j}$ consist of the real-space displacement vector and chemical species of neighbor $j$, $(\mathbf{r}_{ij}; s_j)$, where $\mathbf{r}_{ij} = \mathbf{r}_j - \mathbf{r}_i$.
While classification algorithms are often sensitive to such a choice for small neighborhoods, the large neighborhoods used here should be less sensitive since they already include multiple coordination shells.
Note that despite using an isotropic cutoff, the inclusion of bond angle features allows for the explicit treatment of anisotropic structures.
From this $n_i$ nearby particles, a collection of rotation-invariant, three-body terms (or features, $\mathbf{F}_i$) between particles $(i, j, k)$ is computed:
\begin{itemize}
\item $d_{jk} = |\mathbf{r}_k - \mathbf{r}_j|$ is the distance between neighbors $(j, k)$
\item $\theta_{jik} = \arccos \left( \mathbf{r}_{ik} \cdot \mathbf{r}_{ij} \right)$ is the bond angle
\item $\ell_{jik} = d_{ij} + d_{ik}$ is the bond length
\end{itemize}
Note that particle $i$ necessarily participates in every bond so the total number of features is only $n_i^2$.

Permutation invariance is incorporated by performing a Gaussian expansion in two dimensions for each of the $(d, \theta)$, $(d, \ell)$, and $(\theta, \ell)$ slices.
When multiple species are present, multiple histograms $\mathbf{H}_i^s$ are generated, with the three-body terms being included only if $j$ or $k$ belongs to species $s$.
For instance, given species $A$ and $B$, I generate two histograms $\mathbf{H}_i^A$ and $\mathbf{H}_i^B$.
The histogram $\mathbf{H}_i^A$ includes all $(i, j, k)$ where either $j$ or $k$ is species $A$; the identify of $i$ is not considered because it is the same for every bond.
Thus $\mathbf{H}_i^A$ is constructed from $AA$ and $AB$ pairs, while $\mathbf{H}_i^B$ is constructed from $BB$ and $BA$ pairs.
This choice does potentially reduce the discriminative power of the method by including cross-terms in multiple histograms, but it ensures a linear scaling with the number of species (e.g., for 5 species only 5 histograms are required instead of the 15 unique combinations).

The histogram $\mathbf{H}_i$ is also normalized to avoid a strong dependence on the number of particles, an unfortunate feature I observed in my graph-based approach \cite{Reinhart2018Automated}.
The resulting tensor has $n_\mathrm{species} \times 3 \times n_\mathrm{bins}^2$ rotation-invariant features.
The choice of two-dimensional slices for $\mathbf{H}_i$ instead of the full three-dimensional tensor is motivated by two related factors.
The $n_\mathrm{bins}^3$ number of Gaussians required for the full 3D case was quite high, and both computational time and data storage are negatively impacted.
In addition, the histograms become more sparse, and the bin width had to be larger to achieve a reliable distance metric (see Section~\ref{sec:umap}).
I found that similar environments (e.g., HCP and FCC crystals) could still be distinguished in the 2D case, while comparisons between dissimilar environments was more reliable.
However, I expect there are cases where the cost of retaining the full 3D information in $\mathbf{H}_i$ is worthwhile, and nothing in the workflow which follows precludes this option.

I note that the computational burden of this scheme is considerably higher (perhaps $10 - 100 \times$) than other common representations whose features are selected by hand to yield physically meaningful information.
For example, Common Neighbor Analysis\cite{Honeycutt1987} computes only three features for each atomic neighborhood, but each feature is known to be highly discriminative.
Likewise, Bond Order Analysis\cite{Steinhardt1983, Lechner2008} was designed to capture symmetries in the atomic environment which are known to have a close relationship to crystal space groups.
Here, I compute many more features than are (probably) necessary because there is an obvious scheme to do so rather than because there is an obvious use for them.

\subsection{Manifold learning}\label{sec:umap}

To obtain a low-dimensional embedding, $\mathbf{H}_i$ is reshaped into a $(3 \, n_\mathrm{species} \, n_\mathrm{bins}^2) \times 1$ feature vector.
A low-dimensional manifold is obtained from embedding $n_\mathrm{obs}$ such feature vectors using the UMAP algorithm \cite{McInnes2018UMAP}.
UMAP is an unsupervised dimensionality reduction algorithm that assumes the data is uniformly distributed on a Riemannian manifold, the manifold is locally connected, and the Riemannian metric on the manifold is locally constant.
This results in a low-dimensional manifold (i.e., projection) which approximates the fuzzy topological structure of the high-dimensional data.
I refer to the low-dimensional manifold as $\lambda$ throughout this paper.

UMAP has three primary hyperparameters: dimensionality of the manifold, local connectivity of the data, and minimum distance between observations in the manifold.
For embedding local environments, I fix the dimensionality to 3 based on prior experience with manifold learning for these systems and ease of visualization.
The manifold topology was relatively insensitive to the number of neighbors since the environments varied smoothly, so the resulting global structure of the manifold remained intact even with low coordination.
I set the minimum distance to be zero because ground state crystal lattices should necessarily have zero distance between their environments; enforcing a finite distance in the manifold would cause distortion to the manifold topology around these ground states due to the artificial enforcement of a maximum density.

\subsection{Collective variables}\label{sec:CV}

Once the individual atomic environments have been embedded in the low-dimensional manifold, the same procedure can be repeated to obtain embeddings of collections of environments such as an entire crystalline domain, simulation box, or simulation trajectory.
In exactly the same way as the bonds in individual environments, the histogram approach described above imparts permutation invariance to this collection of environments.
To obtain collective variables for the entire simulation domain, I therefore generate a histogram $\mathbf{H}^\lambda$ (i.e., Gaussian expansion) of the $\lambda_i$ for all $i$ in the simulation snapshot.
Then the histogram features can be embedded in a second low-dimensional manifold $\xi$ which describes the character of the entire snapshot.
Each dimension of $\xi$ can then be treated as a collective variable.
For these collective variables I use different UMAP hyperparameters since the size and physical meaning of the data set is substantially different.
First, I reduce the manifold dimension of 2 for simpler visualization and under the premise that variation in fewer dimensions will correspond to broader trends in the data.
I set the minimum distance to a small but nonzero value (0.05) to enforce variation between simulation frames; this was an empirical choice based on an improvement in the interpretability of the results.
Finally, I increase the connectivity to include half the data set to enforce smoother topological structuring -- for the individual atomic environments there are many thousands of observations available which leads to a smooth manifold, but with only tens of snapshots the manifolds are prone to spurious local ordering.

\subsection{Classification}\label{sec:classification}

Supervised classification is used in the classification of ices and binary mesophases to compare directly to the results of PointNET from Ref.~\cite{Defever2019}.
I use a Random Forest Classifier (RFC) as implemented in the popular scikit-learn software package \cite{scikit-learn}.
The models use a max decision tree depth of 10 and ensemble size of 100 decision trees.
The tree depth is limited to avoid learning over-fitting, especially because the low-dimensional manifolds offer a relatively simple topology which can lead to spurious interpretation with a low volume of training data.
Additional details concerning the training data are given on a case-by-case basis in Sections~\ref{sec:ice} and \ref{sec:mesophase}.

\section{Results and discussion}

\subsection{Colloidal crystallization}\label{sec:colloids}

I begin with an analysis of the same evaporation-induced colloidal crystallization problem I have studied in previous articles \cite{Reinhart2017Machine,Reinhart2018Automated}.
The details of the Molecular Dynamics simulations used to generate these data are available in Ref.~\cite{Howard2018Evaporation}.
This system makes for a good case study because it exhibits several features which are challenging for classification algorithms:
(i) there are few particles in the system so the crystal domains are relatively small,
(ii) it is out of equilibrium so some transient, metastable states appear,
(iii) there is a sharp interface between the vapor and condensed phases which locally disrupts the coordination,
(iv) the attractive interactions between particles are weak which results in large thermal noise.
As in Refs.~\cite{Reinhart2017Machine,Reinhart2018Automated}, I analyze 50 snapshots with $2\,052$ particles each.

\begin{figure*}[!t]
\center
\includegraphics[width=0.9\textwidth]{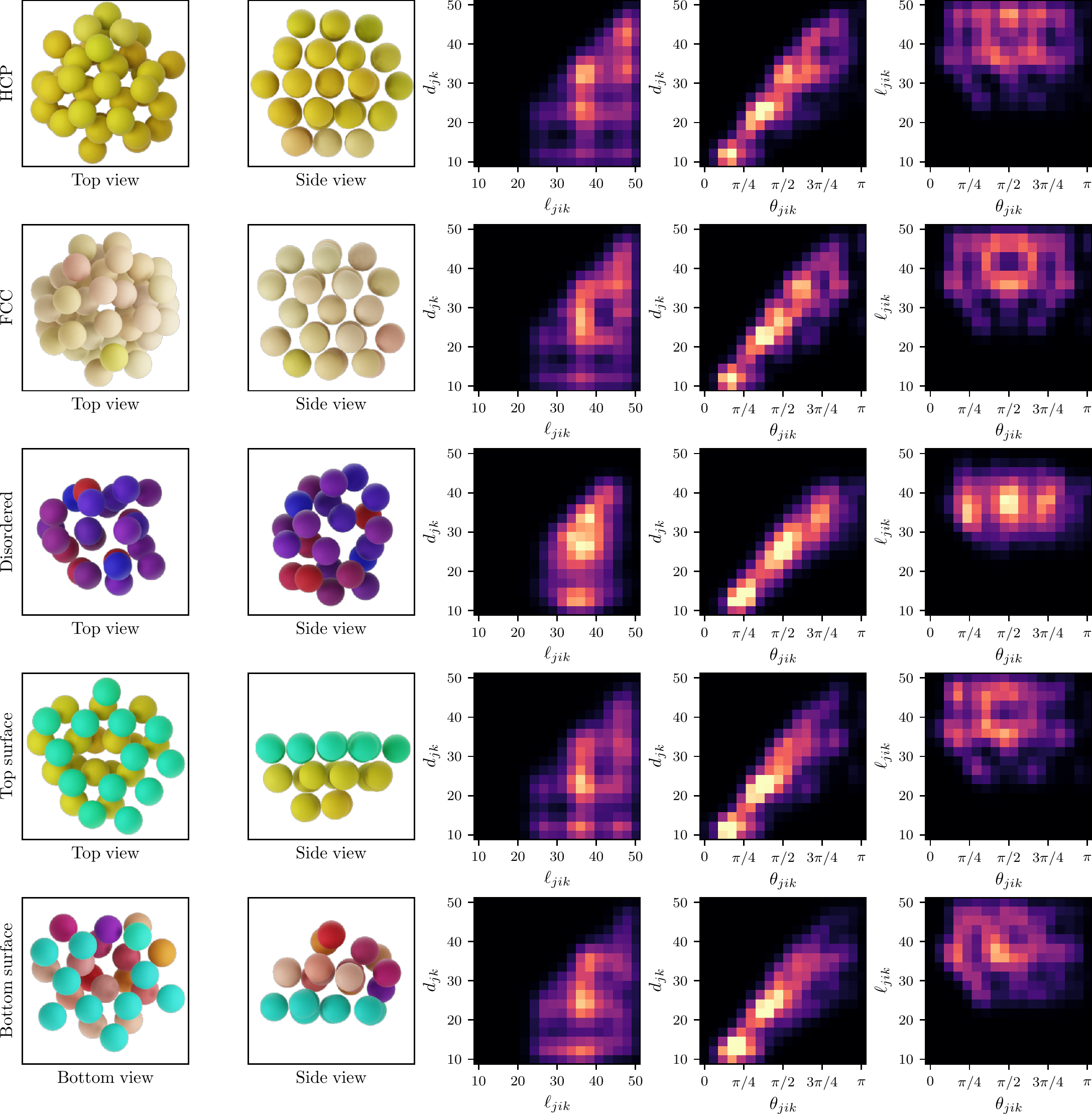}
\caption{
Renderings of the real-space neighborhood, $\mathbf{R}$, and two-dimensional projections of the three-dimensional histogram, $\mathbf{H}$, for five key structures.
Particle $i$ is not rendered since it is implicit in $\mathbf{R}_i$ (i.e., particle $i$ always appears at the origin).
Lighter values have higher probability density in the histograms.
Particles are colored according to their position in $\lambda$-space, as illustrated in Fig.~\ref{fig:colloids-umap}.
\label{fig:colloids-histograms}
}
\end{figure*}

Representative histograms $\mathbf{H}$ are shown alongside real-space neighborhoods $\mathbf{R}$ for five key structures in Fig.~\ref{fig:colloids-histograms}.
The distinction between HCP and FCC crystal structures (top two rows) can be clearly seen in the $(\theta_{jik}, \ell_{jik})$ slice (rightmost histogram), and more subtly in the $(\ell_{jik}, d_{jk})$ slice (leftmost histogram).
Notably, the clearest distinction appears at $\ell_{jik} \approx 45$, which corresponds to the second neighbor shell.
The disordered phase (middle row) looks qualitatively different than the crystal phases, with less distinct ordering visible in the histogram.
There is also a narrower range of $\ell_{jik}$ visible, presumably because the lower local density in the disordered region shifts the coordination shells farther away and excludes the second shell from the fixed cutoff radius.
The surface environments (bottom two rows) are somewhat in between the crystal and disordered phases, with ordering visible but subdued compared to the crystal phases.
In these cases, there is an asymmetry in the $\theta_{jik}$ coordinate of the $(\theta_{jik}, \ell_{jik})$ slice (rightmost histogram) which is not present in the other three cases.
Specifically, there is a bright spot near $(\pi / 3, 35)$ in both cases which is not mirrored by a similar spot at $(2 \pi / 3, 35)$, contrasting the symmetry across $\theta_{jik}$ in the equivalent slice for HCP and FCC.

\begin{figure}[!t]
\center
\includegraphics{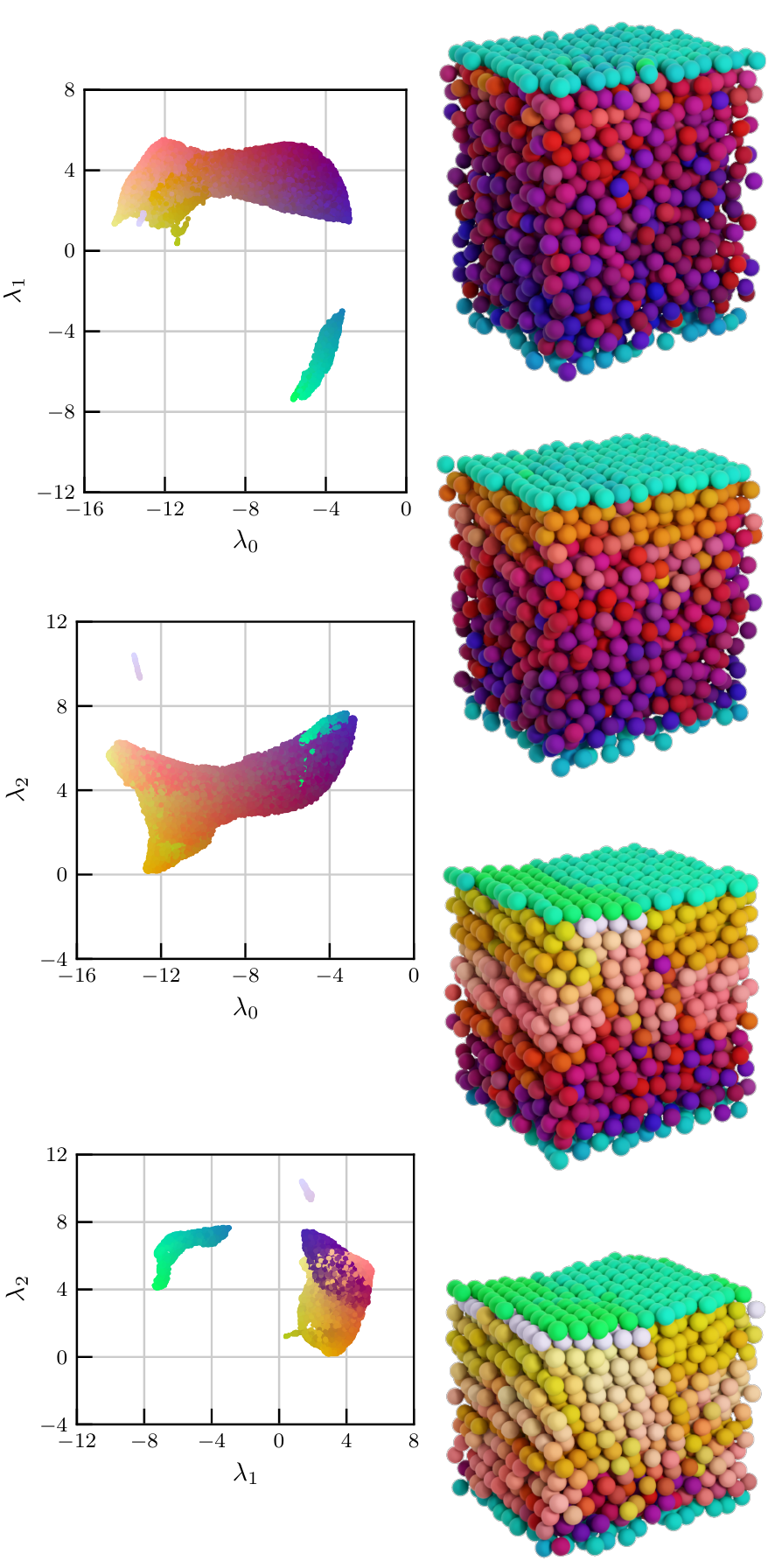}
\caption{
(Left) Three two-dimensional projections of the three-dimensional UMAP for evaporation-induced colloidal crystallization.
(Right) Snapshots of the colloidal system crystallizing during evaporation with particles colored according to their position in $\lambda$-space shown in the top panel.
\label{fig:colloids-umap}
}
\end{figure}

The three-dimensional UMAP obtained from embedding the $2\,052 \times 50$ individual histograms is shown in Fig.~\ref{fig:colloids-umap}.
While the latent variables are never guaranteed correspond to intuitive quantities, the overall structure of the manifold does typically provide some insight into the relationships between the observed environments.
In this case, the $\lambda_0$ dimension appears related to coordination number, with bulk crystalline environments appearing at low $\lambda_0$ while disordered and surface crystalline environments appear at high $\lambda_0$.
This is corroborated by the observation from Fig.~\ref{fig:colloids-histograms} regarding the exclusion of a second neighbor shell in the disordered state.
Furthermore, the $\lambda_1$ dimension appears to distinguish between surface and bulk environments, with the top and bottom interfaces at low $\lambda_1$ clearly separated from the rest of the environments at high $\lambda_1$.
Again, this was clearly observed in the anisotropy of the surface-like histograms in Fig.~\ref{fig:colloids-histograms}.

Finally, the $\lambda_2$ dimension appears related to the crystalline symmetry of the environment, with FCC-like structures ($ABC$ stacking) at high $\lambda_2$ and HCP-like environments at low $\lambda_2$ ($ABA$ stacking).
This might even explain the location of the liquid environments at higher $\lambda_2$ compared to the HCP crystal, and the location of the surface-like environments along the upper-$\lambda_2$ edge of the main cluster.
The anisotropy of the surface-like structures is obvious, since they include a sharp interface between bulk and vacuum.
However, liquids may also belong to this group due to the distinction between \textit{radial isotropy}, which is typical of liquids, and \textit{angular symmetry}, which liquids do not possess; indeed, liquids are already often identified through bond order parameters which consider angular symmetry of specific orders.
It is important to note that these are subject to change with different choices of hyperparameters in the generation of $\mathbf{R}$, $\mathbf{H}$, or $\lambda$.

Some representative snapshots from different times in the trajectory are shown in the bottom panel of Fig.~\ref{fig:colloids-umap}.
The colloids are colored according to their position in the UMAP shown above.
A gradient from purple/blue to red/orange and finally yellow/white is visible from bottom to top of each snapshot (due to heterogeneous nucleation at the top) and left to right across snapshots as the crystallization progresses.
Polymorphic growth is visible in both the surface (light green versus teal) and in the bulk (pink/white versus yellow/orange).
Note that the manifold is smooth in $\lambda$ space, so the classification of particles varies continuously in between the recognizable phases at the corners.
This may have important implications for identifying structure in phase transitions which do not have a well-known local order parameter, such as those exhibited by dynamical systems.\cite{Royall2020}

\begin{figure}[!t]
\center
\includegraphics[width=9cm]{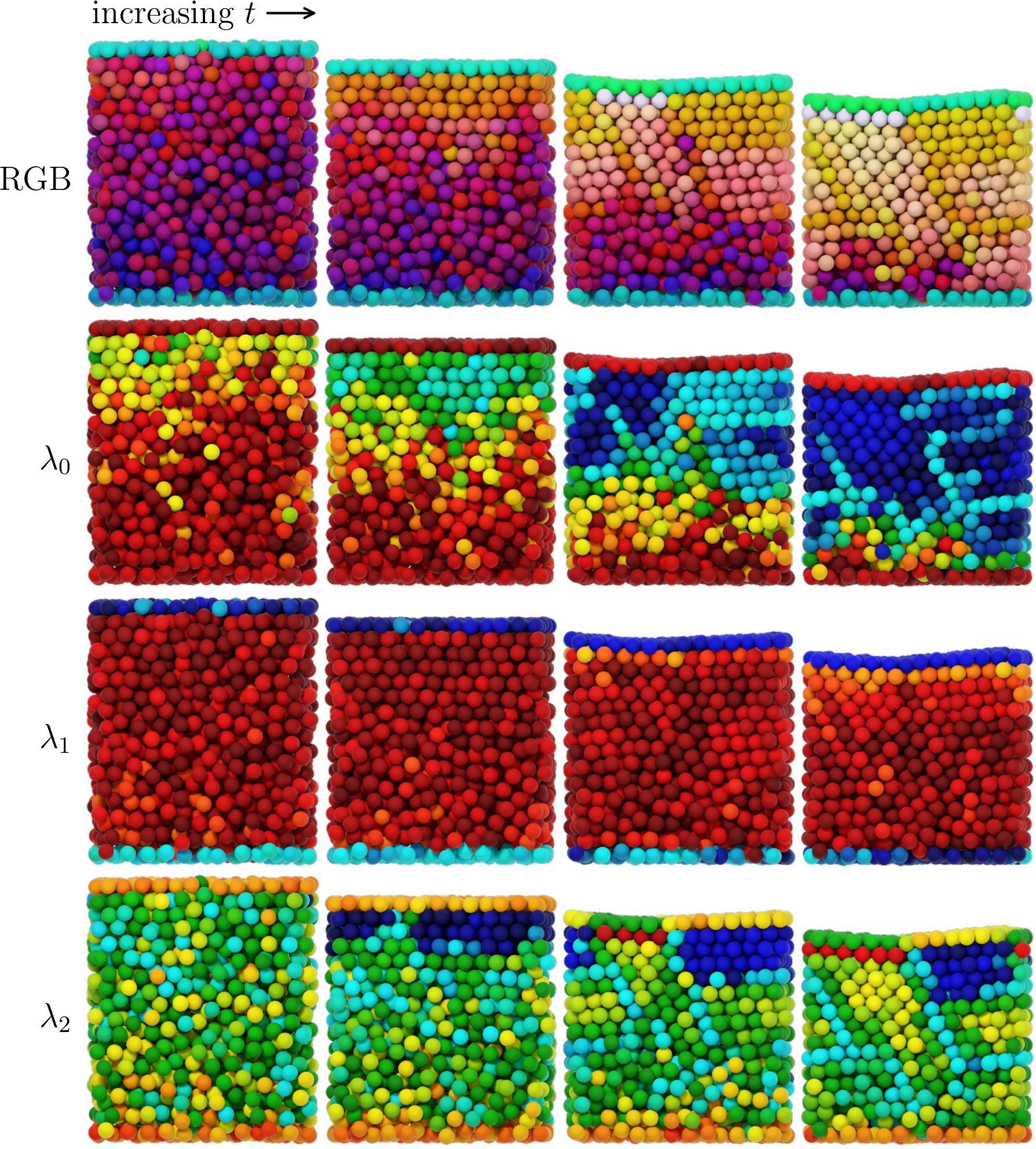}
\caption{
Same snapshots as in Fig.~\ref{fig:colloids-umap} colored according to their position in each of the three latent dimensions.
Color progression is red--yellow--green--cyan--blue.
\label{fig:colloids-lambdas}
}
\end{figure}

Of course, the value of such visualizations is subject to human perception and it is important to point out that while the distinction between FCC and HCP appears weak in this color scheme, it is strongly reflected in the values of $\lambda_2$.
I therefore reproduce the same snapshots using three additional color schemes based on the individual $\lambda_i$ values in Fig.~\ref{fig:colloids-lambdas}.
In those snapshots, one can clearly observe the progress of the crystallization in $\lambda_0$, with dark blue corresponding to both FCC and HCP domains with light blue appearing at both early stages and later at stacking faults.
On the other hand, $\lambda_1$ barely changes over the course of the simulation, with the surfaces clearly distinguished from the bulk structures throughout.
Interestingly, the light blue bottom surface finally transitions to dark blue at the end of the trajectory when the crystal domain extends all the way through to the bottom interface.
Finally, $\lambda_2$ indicates symmetry, so the dark blue appears in the HCP domain, with yellow in the FCC domain, light blue at the grain boundaries.
The single red layer corresponds to a stacking fault seemingly related to competition between surface tension and the geometric constraints encountered by the bulk crystal, since it appears only after the surface buckles in the third snapshot.

\begin{figure}[!t]
\center
\includegraphics[width=7cm]{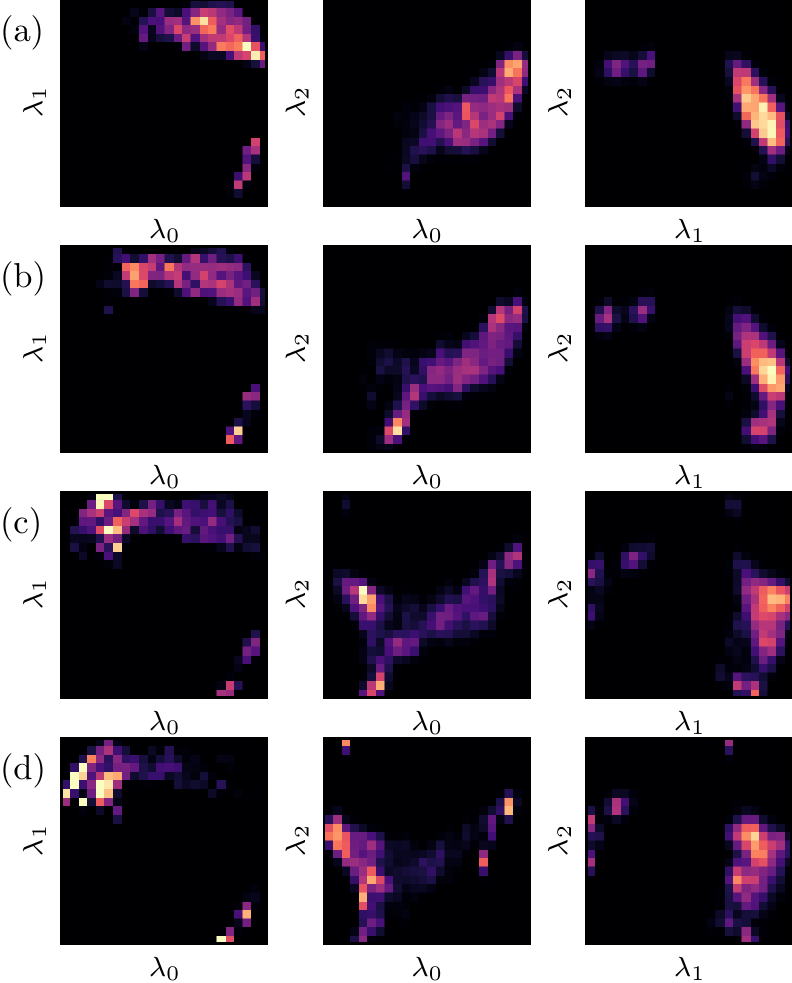}
\caption{
Examples of $\lambda$-space histograms, $\mathbf{H}_\lambda$, for the colloidal system.
Rows correspond to the same snapshots shown in Fig.~\ref{fig:colloids-umap}: (a) $t = 15$, (b) $t = 25$, (c) $t=35$, (d) $t=45$.
\label{fig:colloids-snapshot-histograms}
}
\end{figure}

In addition to classifying individual particle environments, the UMAP embedding can be used to construct collective variables for the entire system.
In this test case, the constant evaporation rate creates a clear arrow of time which should be straightforward to detect.
I compute the collective variables by generating another Gaussian expansion histogram, but this time based on the positions of each particle in the snapshot in $\lambda$-space, as described in Section~\ref{sec:CV}.
Thus, a histogram $\mathbf{H}^\lambda$ is computed from each simulation snapshot.
Representative $\mathbf{H}^\lambda$ are shown in Fig.~\ref{fig:colloids-snapshot-histograms}.
Then these system-wide histograms can be flattened into feature vectors and embedded in a second UMAP in nearly the same way the individual particle environments were.

\begin{figure}[!t]
\center
\includegraphics[width=9cm]{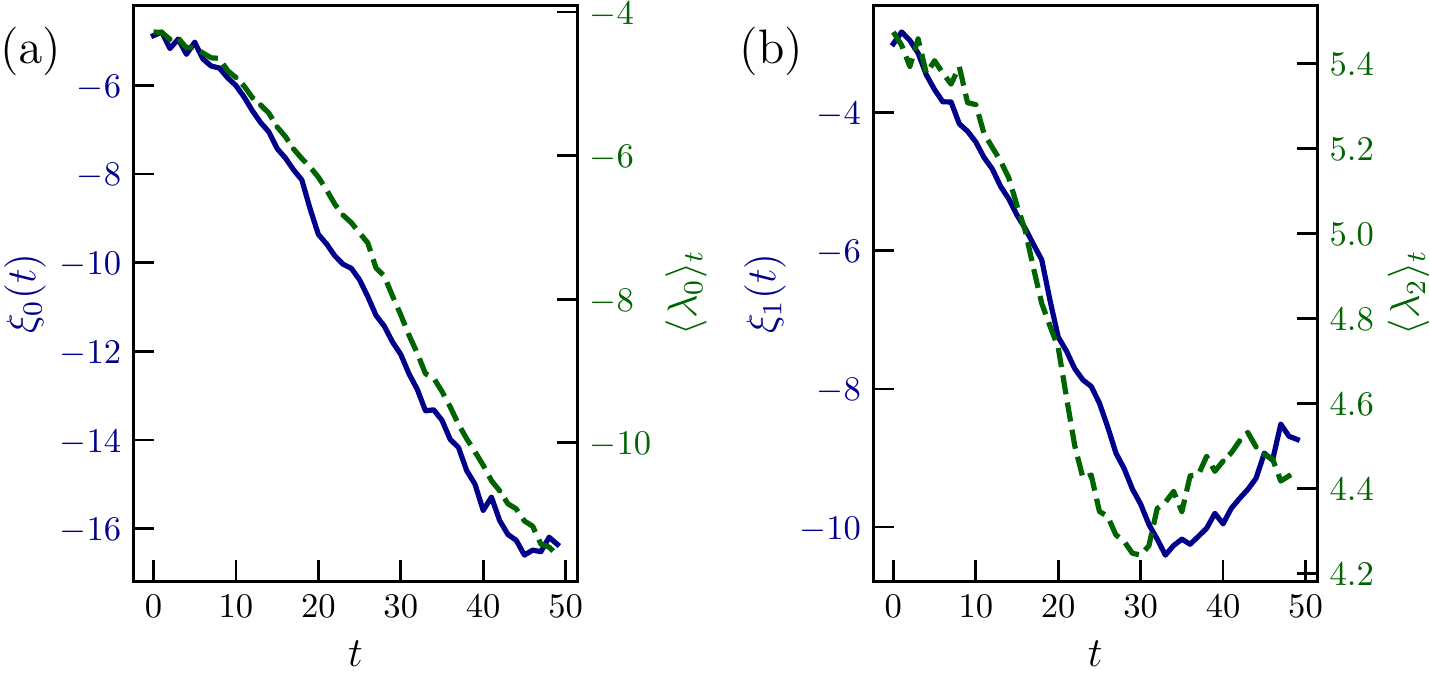}
\caption{
Collective variables from $\mathbf{H}_\lambda$ UMAP for colloidal crystals.
(a) The first coordinate $\xi_0$ (left axis, blue solid line) and average $\lambda_0$ (right axis, green dashed line) from each frame plotted against the simulation time $t$ (arbitrary units).
(b) The second coordinate $\xi_1$ (left axis, blue solid line) and average $\lambda_2$ (right axis, green dashed line) from each frame against $t$.
\label{fig:colloids-rxn-coords}
}
\end{figure}

The two collective variables of the system-wide UMAP, which I refer to as $\xi$, are shown in Fig.~\ref{fig:colloids-rxn-coords}.
The hyperparameters used in this embedding were selected to yield smooth curves in the time trace; when choosing a small number of neighbors or a very small minimum distance, the topology of the manifold can become disconnected and lead to discontinuous $\xi(t)$.
As such they are quite subjective, but I also note that the general structure of $\xi$ is not strongly sensitive to the choice of hyperparameters above a certain threshold.
Plotted on separate axes are some hand-chosen features which share the same qualitative features -- these are the average value of $\lambda_0$ and $\lambda_2$ in each snapshot.
Note that these were selected by hand to demonstrate how physical meaning can be embedded in the results obtained by unsupervised learning, even though the numerical values are the result of a `black box' (the embedding scheme).
Again, the comparisons made here merely illustrate the potential insights which might be gained from the unsupervised generation of collective variables $\xi$, since the particular features of the UMAP manifold are subject to the choice of hyperparameters.

The average $\lambda_0$ should roughly correspond to the degree of crystallization, though many other measurements may share a qualitatively similar trend due to the constant evaporation rate.
The nearly constant value of $\xi_0$ for $t < 10$ is a result of the delayed onset of heterogeneous nucleation due to insufficient local density in the early stages of the simulation \cite{Howard2018Evaporation}.
Here the increase in $\langle \lambda_0 \rangle$ just before $\xi_0$ is probably a result of increasing density which drives the local environments to a different part of $\lambda$-space.
The tapering rate of both measures towards the end of the simulation shows that that a maximum density is reached.

For $\xi_1$, the qualitative trend appears to match $\langle \lambda_2 \rangle$.
A decreasing $\lambda_2$ indicates increasing symmetry associated with migration from disordered to more ordered liquid environments in the early stages, and HCP growth in the later stages.
The decrease in symmetry in later frames might be associated with the formation of stacking faults due to collective rearrangements as the crystal domains are forced to accommodate the finite simulation box size (e.g., light blue faults seen in rows 2 and 4 of Fig.~\ref{fig:colloids-lambdas}).
Of course, there are other physical quantities that may exhibit the same shape as $\xi_1$.
One example might be the rate of crystal growth, which should increase throughout the middle of the trajectory and then taper off at the end.

\subsection{Ice nucleation}\label{sec:ice}

I next consider the heterogeneous nucleation and growth of ice phases, using the same data as in Ref.~\cite{Defever2019}.
In that study, a supervised PointNet model was trained on eight different phases, including liquid, five ice phases, and two hydrate phases.
For bulk phases, they achieved overall accuracy of $>99\%$, which exceeded the results of Geiger and Dellago.
They had to train on both interfacial and bulk phases to get good accuracy in the heterogeneous nucleation case.
To train PointNet, $10^4 - 10^5$ observations of each phase were used, and the model weights were optimized over 100 epochs.
This directly contrasts with the manifold learning procedure used here which requires no training but does take preprocessed features (i.e., $\mathbf{H}$) as inputs rather than raw particle positions (i.e., $\mathbf{R}$).
Therefore the computational cost of my method is minimized for small data sets but grows linearly with the number of observations, while PointNET effectively amortizes the upfront training cost through sub-linear scaling on large data sets.

Here, I consider the center of mass of each water molecule, which is approximately equal to the position of the oxygen atom.
While the proposed methodology supports the use of multiple species and multiple atoms per molecule, the classification is faster with only one point per molecule, and the results are already quite good.
I first obtain a low-dimensional manifold using unsupervised UMAP, following the scheme described in Section~\ref{sec:colloids}.
This manifold contains only $1\,000$ samples of each phase, which is not even one full snapshot of each bulk phase considered in Ref.~\cite{Defever2019} (they range from about $500$ to $2\,000$ water molecules per snapshot).
Herein lies the main advantage of the proposed unsupervised method -- only a very small amount of data is required to obtain a meaningful description of the relationships between environments.
Importantly, there are no data labels included at this stage -- only the raw features are used to construct the manifold.

\begin{figure}
\center
\includegraphics{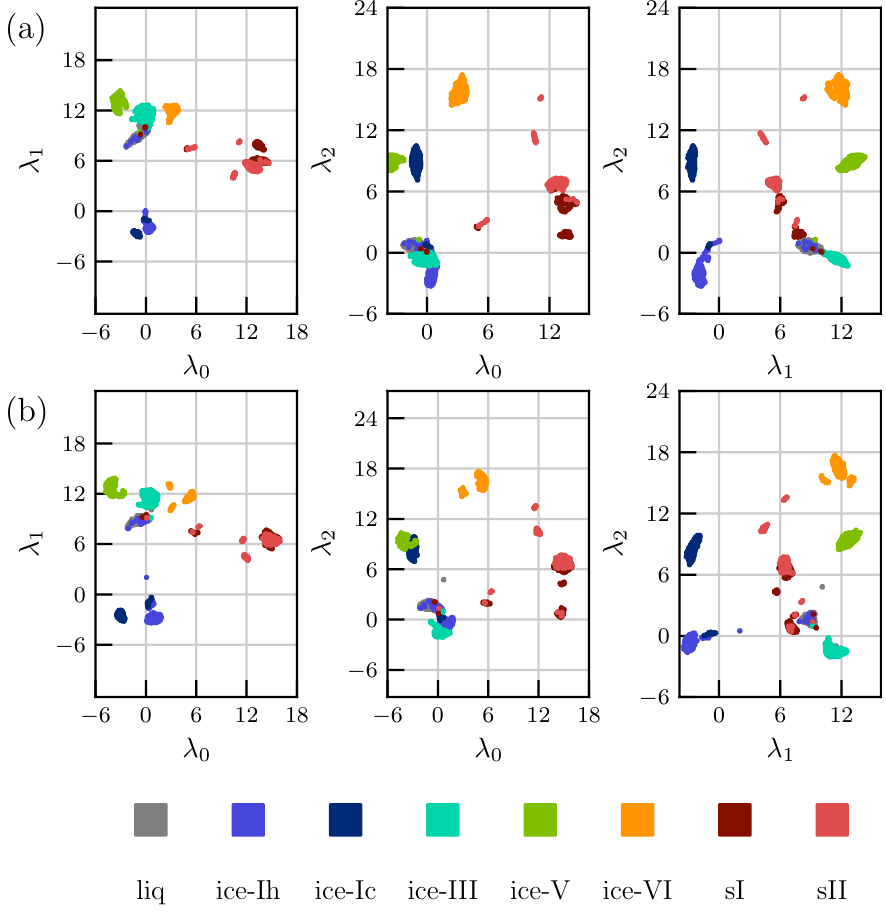}
\caption{
Manifold obtained from unsupervised UMAP on bulk ice structures using categorical color code shown below.
(a) Ground truth labels from $1\,000$ samples of each ice structure in the training data set.
(b) Ground truth labels from $1\,000$ samples of each ice structure in the testing data set.
\label{fig:ice-umap}
}
\end{figure}

The three-dimensional manifold obtained from embedding the $1\,000 \times 8$ bulk ice environments is shown in Fig.~\ref{fig:ice-umap}.
The color code refers to the bulk structure from which each environment was sampled; since these are the ground truth labels the spatial distribution of labels of the same color is a real, physical phenomenon rather than an artifact of an imperfect model.
The snapshots are equilibrated at finite temperature ($T = 200-300 \; \mathrm{K}$) and therefore exhibit large structural fluctuations.
This thermal noise leads to liquids being somewhat obscured in this plot by molecules from snapshots with bulk ordering but local disorder leading to a more liquid-like environment.
The discrete clusters observed for sI/sII are due to true differences in atomic environment resulting from the large unit cell of the hydrate cages.
Panels (a) and (b) correspond to two different samples of $1\,000 \times 8$ environments taken from different snapshots.
The difference in cluster topology (e.g., ice-VI cluster in $\lambda_0 \mbox{-} \lambda_2$ slice) shows the sensitivity to thermal noise when evaluating only a small number of environments.

Here I do not speculate on the physical meaning of the ice UMAP, and instead point out only the main topological features in each dimension.
The $\lambda_0$ dimension appears to separate the cage-like sI, sII, and ice-VI environments from the rest.
A strong separation is observed in $\lambda_1$ between the ice-Ic/Ih environments, the liquid/hydrate environments, and the ice-III/V/VI environments.
Finally, $\lambda_2$ distinguishes ice-Ih from ice-Ic at low $\lambda_1$ and ice-V from ice-III at high $\lambda_1$.
Obtaining clean separation between the ice-like structures using unsupervised manifold learning suggests that the proposed rotation-invariant features are suitable to capture the subtle differences between a rich variety of crystal phases, not just hexagonal and cubic symmetry as observed in the colloidal crystallization case.

While the UMAP provides discrete clusters which are suitable for unsupervised methods like K-Means, DBScan, or Hierarchical Agglomerative Clustering, these approaches will likely lead to the splitting of clusters associated with a single bulk phase such as the hydrates, since they are apparently distinct environments from the perspective of $\mathbf{H}$.
Since I want to make a direct comparison against the PointNET classification from Ref.~\cite{Defever2019}, I train a supervised Random Forest Classifier (RFC) on this embedding and ensure that cluster labels have a one-to-one correspondence between the two models.
The RFC model is trained using only $75\%$ of the environments from the training set, corresponding to $6\,000$ total atomic environments.
The implementation is from the scikit-learn package \cite{scikit-learn}, and hyperparameters are given above in Section~\ref{sec:classification}.
Training the model takes less than one second, and it achieves accuracy of $>99\%$ on the held-out training data.
This result underscores the robustness of the unsupervised learning method to thermal noise encountered at typical simulation conditions.

\begin{figure}[!t]
\center
\includegraphics{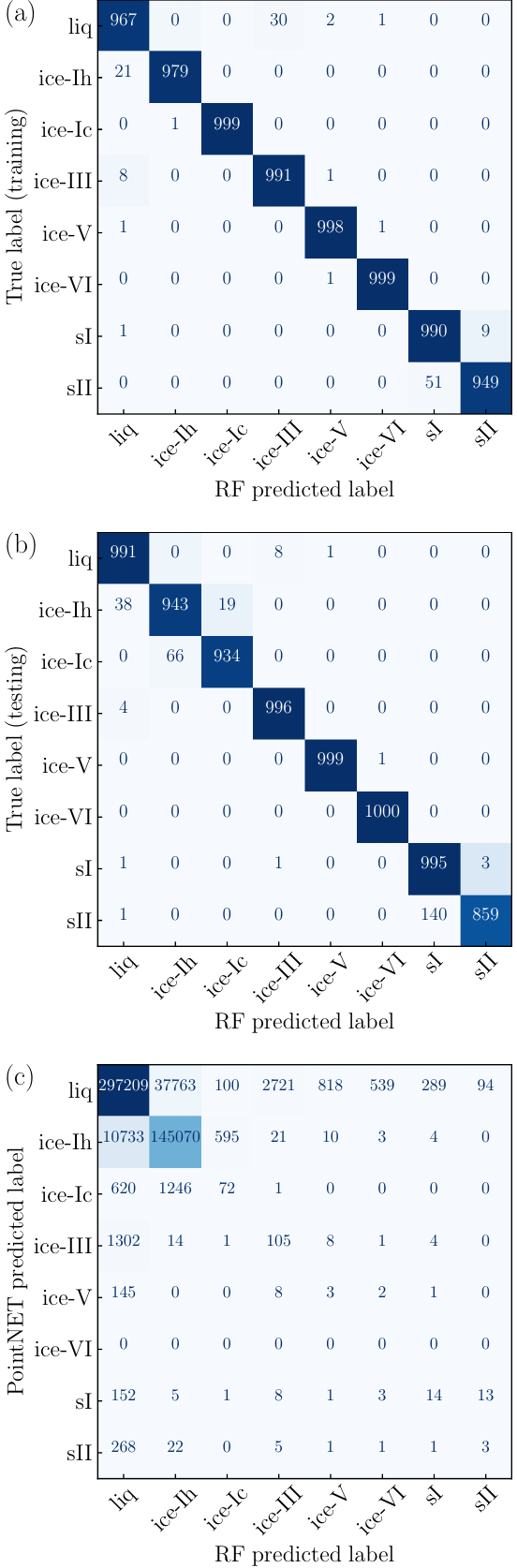}
\caption{
Confusion matrices for Random Forest Clasifier.
(a) Training data set with $1\,000$ samples of each bulk ice structure.
(b) Testing data set with $1\,000$ samples of each bulk ice structure.
(c) Testing data set from nucleation trajectory compared to PointNET result in Ref.~\cite{Defever2019} (ground truth unknown).
\label{fig:ice-confusion}
}
\end{figure}

The RFC model is further tested by taking a new set of $1\,000 \times 8$ bulk ice environments from different snapshots than the training data.
The UMAP for this testing data set is shown in Fig.~\ref{fig:ice-umap}(b).
In this case, the model accuracy reduces to $95\%$, likely due to the topological changes described above.
The confusion matrices for training and testing data sets are shown in Fig.~\ref{fig:ice-confusion}(a) and (b), respectively.
Interestingly, the liquid is actually classified slightly more accurately in the testing case, while ice-Ih, ice-Ic and sII exhibit degraded performance.
This is not surprising given the quite different shapes of the associated clusters from Fig.~\ref{fig:ice-umap}(a) to (b).
Including additional observations -- especially from different snapshots -- in the RFC training set would therefore improve model performance on new data.

This problem further demonstrates the benefits of unsupervised learning compared to supervised learning.
These environments are still located in roughly the same region of the reduced $\lambda$-space, yet the supervised model struggles to identify them because it has learned rules for the boundaries between phases.
Fuzzy classification schemes based on the position in $\lambda$-space (e.g., the color scheme in Fig.~\ref{fig:colloids-umap} or Fig.~\ref{fig:colloids-lambdas}) would not suffer this limitation, but on the other hand they also do not give discrete classes.
So ultimately the choice of discrete or fuzzy classifier both have drawbacks and the better choice depends on the application.

\begin{figure*}[!t]
\center
\includegraphics{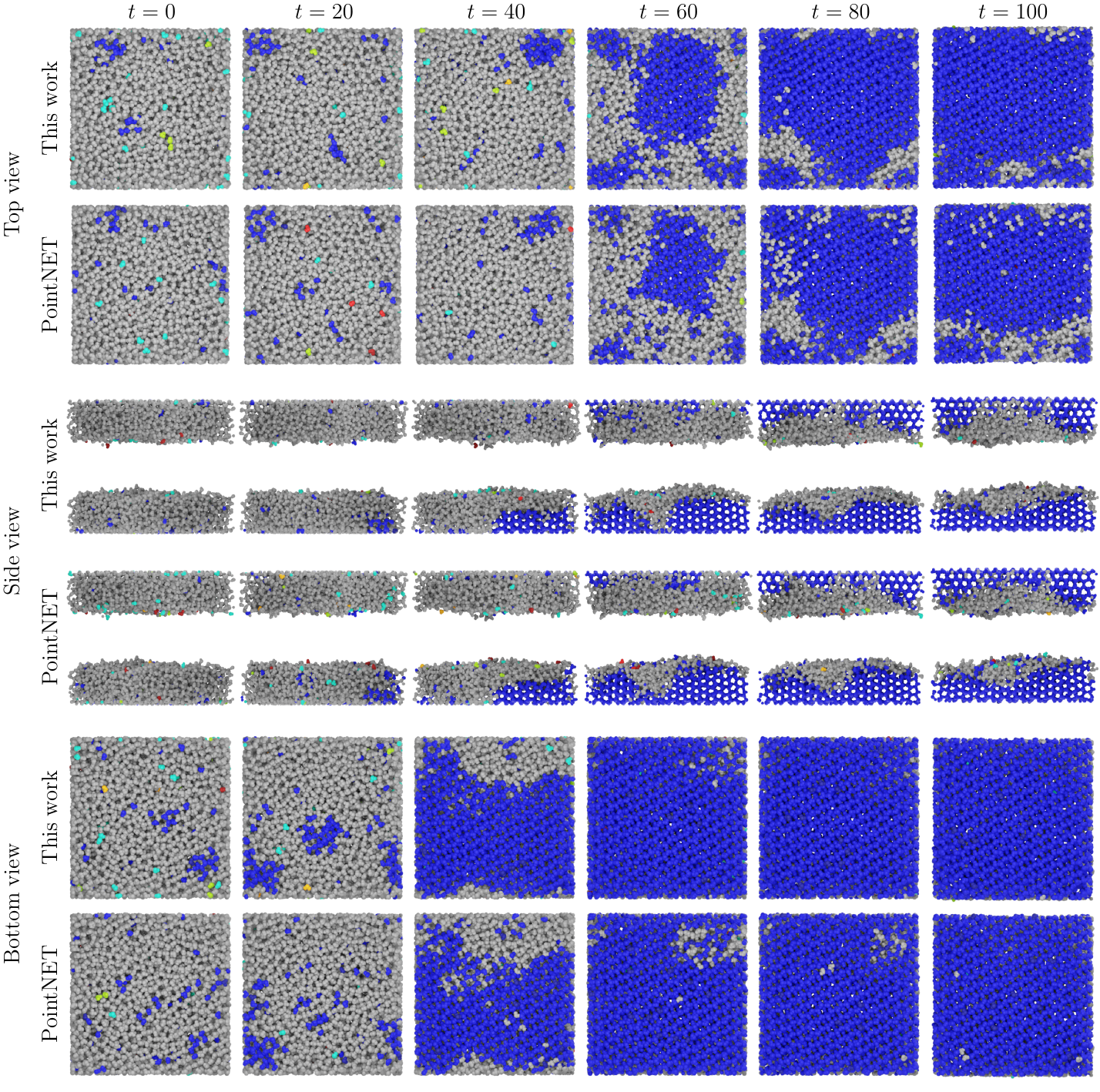}
\caption{
Snapshots with space-filling water molecules colored according to the prediction from the RFC model versus PointNET (categorical color scheme is the same as Fig.~\ref{fig:ice-umap}).
Columns correspond to different times during the simulation, with each pair of rows comparing the result of the RFC and PointNET models for top, side, and bottom views of the same simulation snapshot.
\label{fig:ice-nucleation}
}
\end{figure*}

Snapshots illustrating the performance of the RFC model compared to PointNET are shown in Fig.~\ref{fig:ice-nucleation}.
In this trajectory, ice-Ih crystals nucleate and grow from each of the top and bottom interfaces since the simulation box is not periodic.
Qualitatively, the results are nearly indistinguishable -- both models show crystallites at approximately the same time and location in every snapshot.
Quantitatively, the RFC model assigns about $11\%$ of the molecules classified by the PointNET model as liquid to ice-Ih, while simultaneously assigning a different $3\%$ of those classified by PointNET as ice-Ih to liquid.
These trends can be seen in the confusion matrix in Fig.~\ref{fig:ice-confusion}(c), generated from $100$ frames of the nucleation trajectory.
Both models assign over $98\%$ of the molecules to either liquid or ice-Ih, with the remaining assigned to one of the other structures.
Many of these are likely false positives, although it is entirely possible that some molecules do form transient ice-Ic-like structures, for instance.

Aside from the necessary volume of training data, an important distinction between the approaches is that PointNET was trained on both bulk and interfacial structures in order to capture the `dome'-type environments at interfaces.
On the other hand, the RFC model was trained on only the bulk ice structures, yet it still classifies these interfacial molecules as ice-Ih.
This is likely due to the use of normalized histograms in this work as opposed to a fixed number of points in the PointNET approach; with normalization, deviations in the number of atoms in $\mathbf{H}$ only matters when those deviations introduce qualitatively new structuring.

\begin{figure}[!t]
\center
\includegraphics{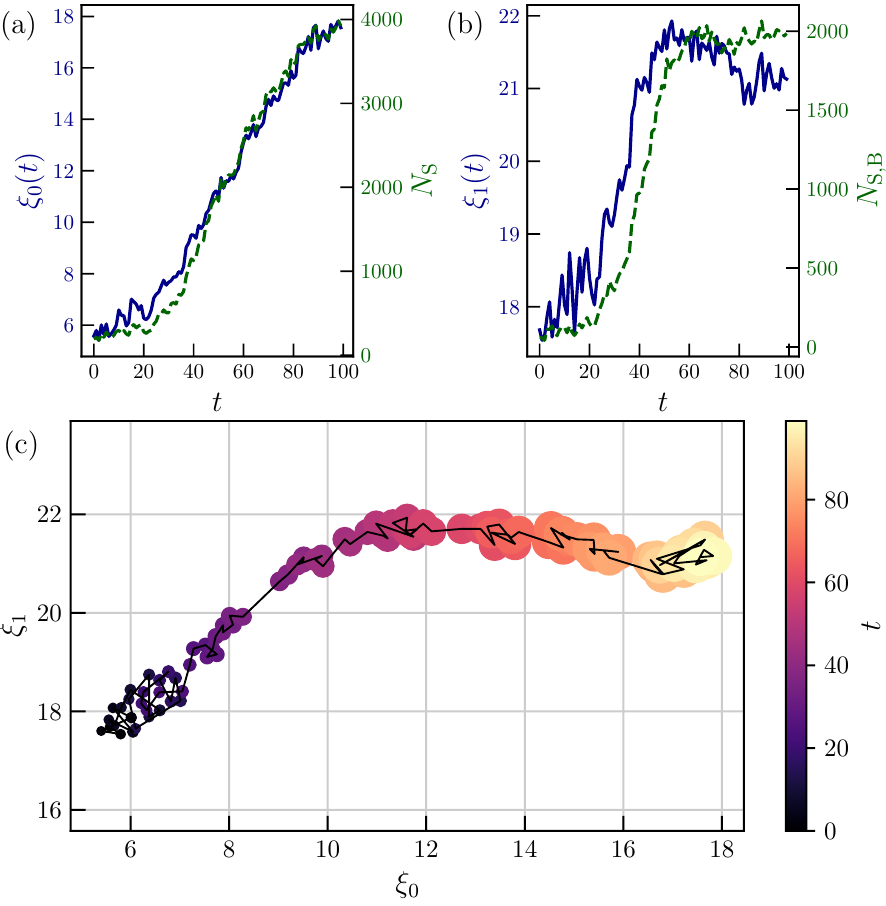}
\caption{
Collective variables from $\mathbf{H}_\lambda$ UMAP for ice structures.
(a) The first coordinate $\xi_0$ (left axis, blue solid line) and number of solid-like molecules (right axis, green dashed line) from each frame plotted against the simulation time $t$ (arbitrary units).
(b) The second coordinate $\xi_1$ (left axis, blue solid line) and number of solid-like molecules near the bottom interface (right axis, green dashed line) from each frame against $t$.
(c) The evolution of $\xi$ throughout the trajectory.
Symbol size corresponds to $N_S$ and symbol color corresponds to $t$.
\label{fig:ice-rxn-coords}
}
\end{figure}

As in the colloidal crystallization test case, I also use the UMAP embedding scheme to construct collective variables for the entire system.
Here the nucleation and growth of ice-Ih crystal domains should delineate the flow of time.
The resulting two coordinates $\xi$ are shown in Fig.~\ref{fig:ice-rxn-coords}.
Plotted on separate axes are some hand-chosen features which share the same qualitative features -- these are the number of solid-like molecules $N_S$ and number of solid-like molecules in the bottom half of the simulation box $N_{S,B}$ in each snapshot.
The $N_S$ in each frame should be strongly reflected in $\xi$ since this trajectory mostly involved molecules transforming from liquid to ice-Ih environments, so the excellent agreement in Fig.~\ref{fig:ice-rxn-coords}(a) is not surprising.

Due to the inclusion of two discrete interfaces in the simulation (shown in Fig.~\ref{fig:ice-nucleation}), it is  reasonable to expect the independent growth of the two crystal domains to appear in the $\xi_1$.
The agreement between $N_{S,B}$ and $\xi_1$ is not perfect, with $\xi_1$ appearing quite noisy both in the beginning and end of the trajectory, and exhibiting a dip away from $N_{S,B}$ towards the end.
Nevertheless, this serves as a good illustration of the sort of qualitative information one can extract from unsupervised manifold learning.
As with the colloidal crystallization case, the comparisons made here are subject to changes in the UMAP topology as hyperparameters are varied.

The trajectory through $\xi$-space is visualized in Fig.~\ref{fig:ice-rxn-coords}(c), with the seemingly random path at low $t$ corresponding to random fluctuations prior to the formation of a critical nucleus.
A relatively straight path up and to the right corresponds the growth of the bottom crystal domain around $t=30$, while the sharp turn to the right corresponds to the growth of the top crystal domain around $t=50$.
At high $t$, the path again wanders randomly around the end point, corresponding to fluctuations in the nearly domain-spanning crystals in both the top and bottom regions.

\subsection{Binary mesophases}\label{sec:mesophase}

\begin{figure*}[!t]
\center
\includegraphics[width=\textwidth]{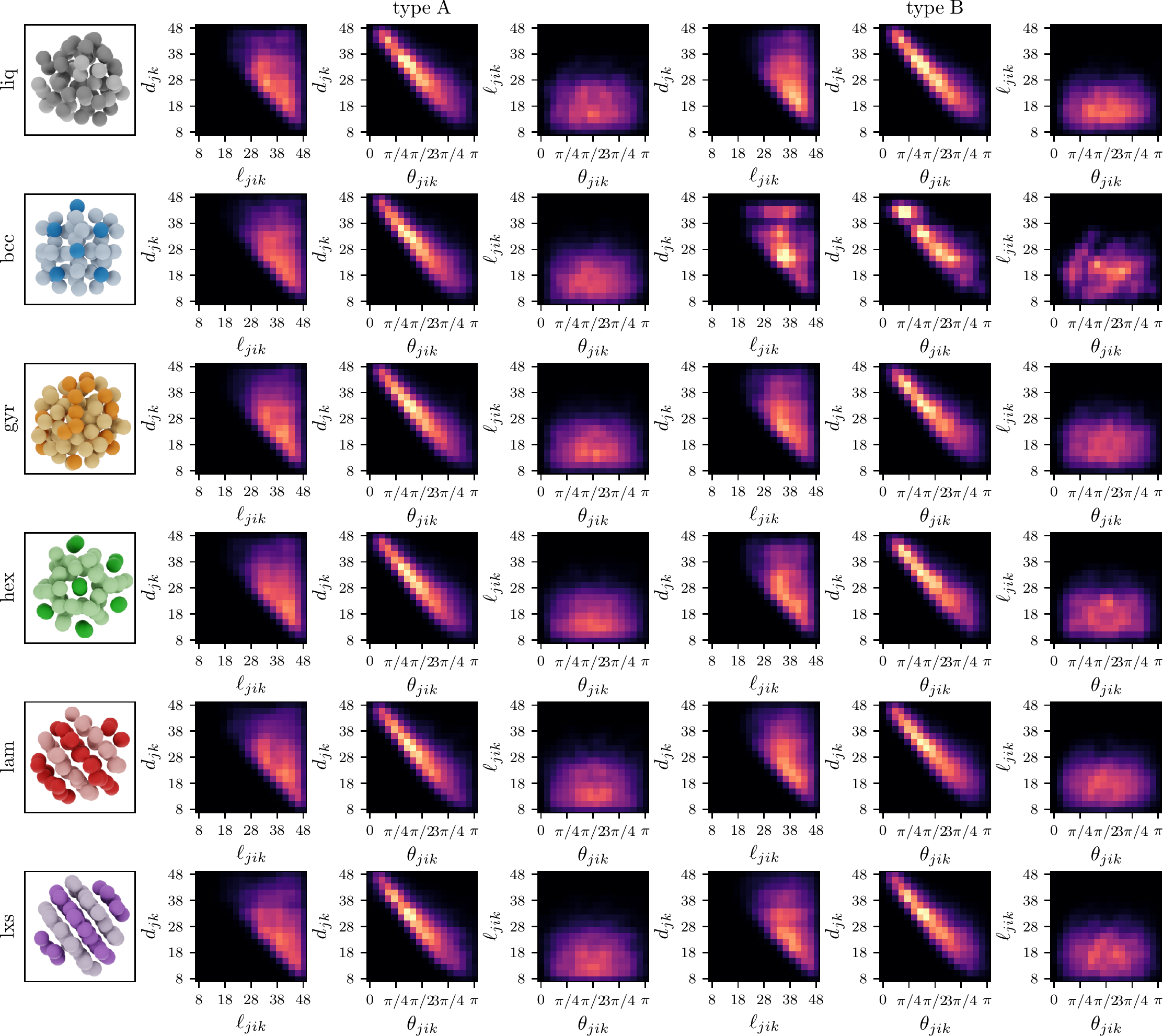}
\caption{
Renderings of the real-space neighborhood, $\mathbf{R}$, and two-dimensional projections of the three-dimensional histogram for each atom type, $\mathbf{H}^A$ and $\mathbf{H}^B$, for the six mesophases.
As in Fig.~\ref{fig:colloids-histograms}, central particle $i$ is not rendered since it is implicit in $\mathbf{R}_i$ (i.e., particle $i$ always appears at the origin).
Lighter values have higher probability density in the histograms.
Particles are colored according to a supervised classifier as described in the text.
\label{fig:meso-histograms}
}
\end{figure*}

I conclude with a case study on binary mesophases \cite{Kumar2017SelfAssembly, Mukhtyar2018Application} to demonstrate extensibility to multiple chemical species.
Binary systems in general pose a serious challenge due to their combined translational and compositional order \cite{Reinhart2017Multi}.
Because particles remain locally amorphous and primarily adopt compositional ordering over multiple coordination shells, order parameters are often ported imprecisely to new problems to provide simple descriptions of order-disorder transitions \cite{Kumar2018Gyroid}, or designed \textit{ad hoc} for a particular study.\cite{Mukhtyar2018Method}
The data presented here are taken directly from Ref.~\cite{Defever2019}, so I follow the same scheme and consider disordered liquid, body-centered cubic (bcc), gyroid (gyr), hexagonal (hex), lamellar (lam), and stacked crystalline phases (lxs).
Two particle species, referred to generically as $A$- and $B$-type, are present in the system.

\begin{figure}
\center
\includegraphics{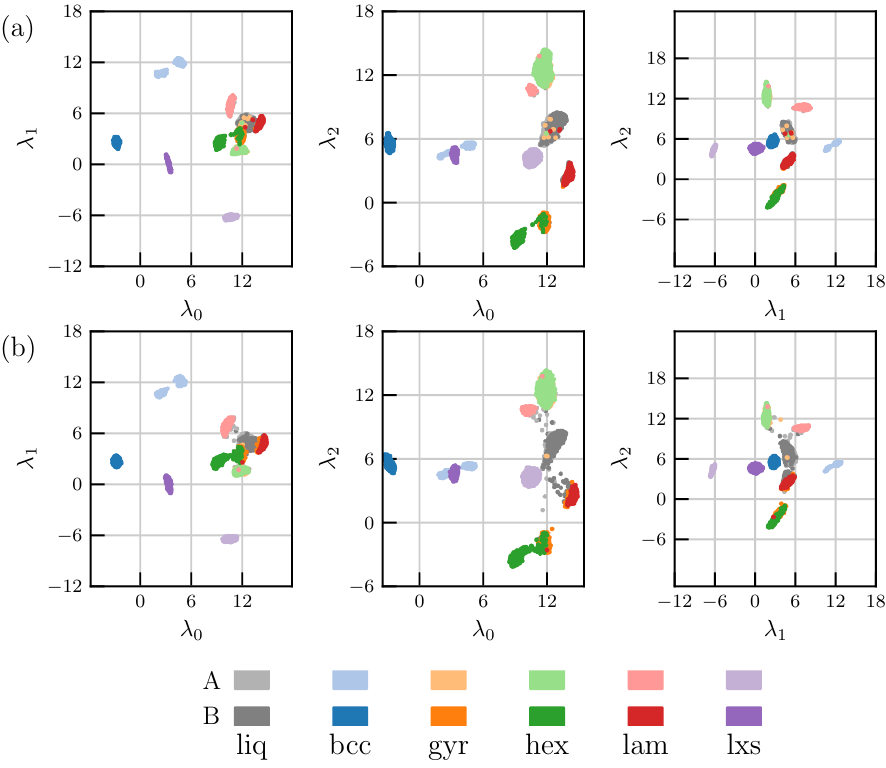}
\caption{
Manifold obtained from unsupervised UMAP on bulk mesophases using categorical color code shown below.
(a) Ground truth labels from $1\,000$ samples of each mesophase in the training data set.
(b) Ground truth labels from $1\,000$ samples of each mesophase in the testing data set.
\label{fig:meso-umap}
}
\end{figure}

Representative $\mathbf{H}$ are shown alongside real-space $\mathbf{R}$ for six mesophases in Fig.~\ref{fig:meso-histograms}.
Among these phases, the bcc shows by far the most distinct features, with ordering reminiscent of the crystalline structures in Fig.~\ref{fig:colloids-histograms}.
The others share generally the same shape with indistinct features due to the high degree of disorder, especially for the $A$-type particles (lighter colors).
Close inspection of these histograms reveals very slight ordering which human intuition can only evaluate qualitatively but which the UMAP algorithm can use to produce a quantitatively meaningful topological structure.

As in the case of the ice structures, the three-dimensional manifold shown in Fig.~\ref{fig:meso-umap} is obtained from embedding $1\,000$ environments from each bulk mesophase, for a total of $6\,000$ environments.
For every mesophase, the environments are split evenly between $500$ $A$-type particles and $500$ $B$-type particles.
The color code again refers to the bulk structure from which each environment was sampled, except now there is a lighter color for $A$ and darker color for $B$, and again these snapshots are equilibrated at finite temperature ($T = 300 \; \mathrm{K}$) and therefore exhibit large structural fluctuations.
The topological structure of the manifold reveals quite distinct structuring for bcc and lxs phases compared to the others -- this is not surprising because these two are crystalline.
From the remaining phases, hexagonal and gyroid $B$-type particles are distinct from many of the others, while the rest are aggregated into a large, central cluster with distinct branches.
This is a result of the spatial and compositional disorder in the $A$-type particles.

\begin{figure}
\center
\includegraphics{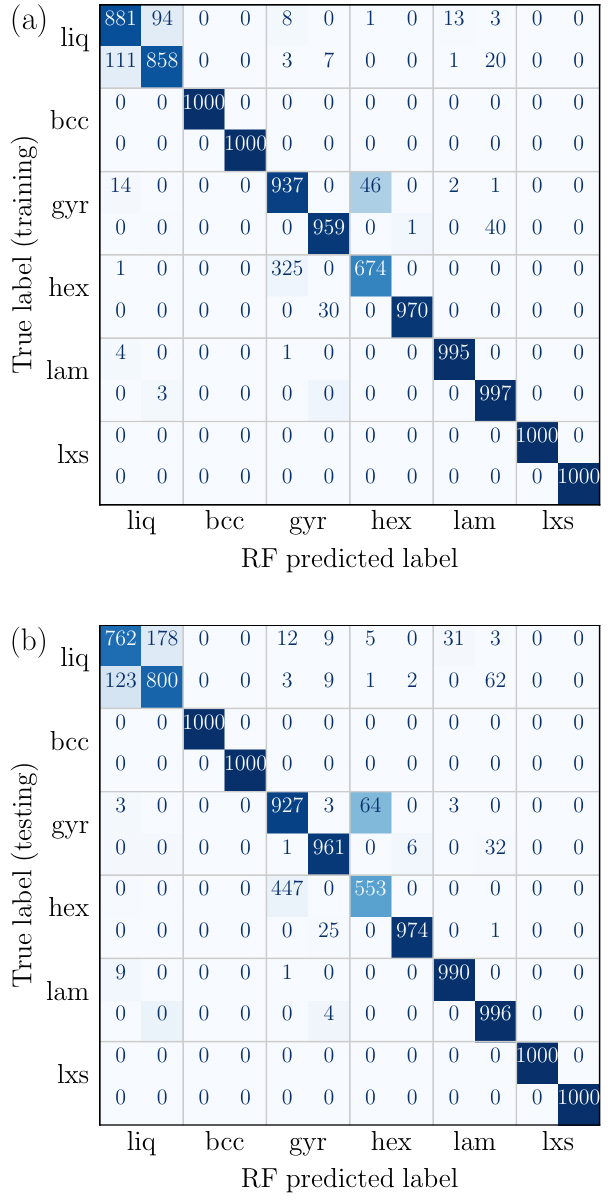}
\caption{
Confusion matrices for the mesophase.
\label{fig:meso-confusion}
}
\end{figure}

To obtain a classification, a RFC model is trained on $2\,000$ environments from one snapshot each of the bulk mesophases.
These are split between $1\,000$ $A$-centered environments and $1\,000$ $B$-centered environments.
The $80\%$ of these environments used for training represent a very small fraction of the available data -- a single snapshot for these mesophases contains between $10\,000$ and $20\,000$ atoms.
On the training data, the RFC achieves $94\%$ accuracy, while on the held-out testing data, this drops to $93\%$.
As with the ices, testing data from different snapshots fairs slightly worse, dropping to $91\%$ accuracy.
The confusion matrices for training and testing data are shown in Fig.~\ref{fig:meso-confusion}(a) and (b), respectively.

Inspecting the confusion matrix in Fig.~\ref{fig:meso-confusion}(b) reveals that a majority of the misclassifications occur between liq-A/liq-B and gyr-A/hex-A environments.
The liquid has no compositional ordering so it is not clear that the disctinction between $A$ and $B$ labels is meaningful in that phase.
Likewise, the $A$ particles in gyroid and hexagonal phases are both in weakly ordered environments, and the distinction may only become meaningful for larger neighborhoods in which the difference between winding and straight columns of $B$ particles is more pronounced.
It is interesting to note that only a few gyr-A are mistaken as hex-A, but nearly half of the hex-A are mistaken for gyr-A (note this is the case even in the training data).
Thus, even in the presence of the highly ordered hex-B rods, the disordered hex-A phase is nearly equivalent to the disordered gyr-A phase.

\begin{figure}
\center
\includegraphics{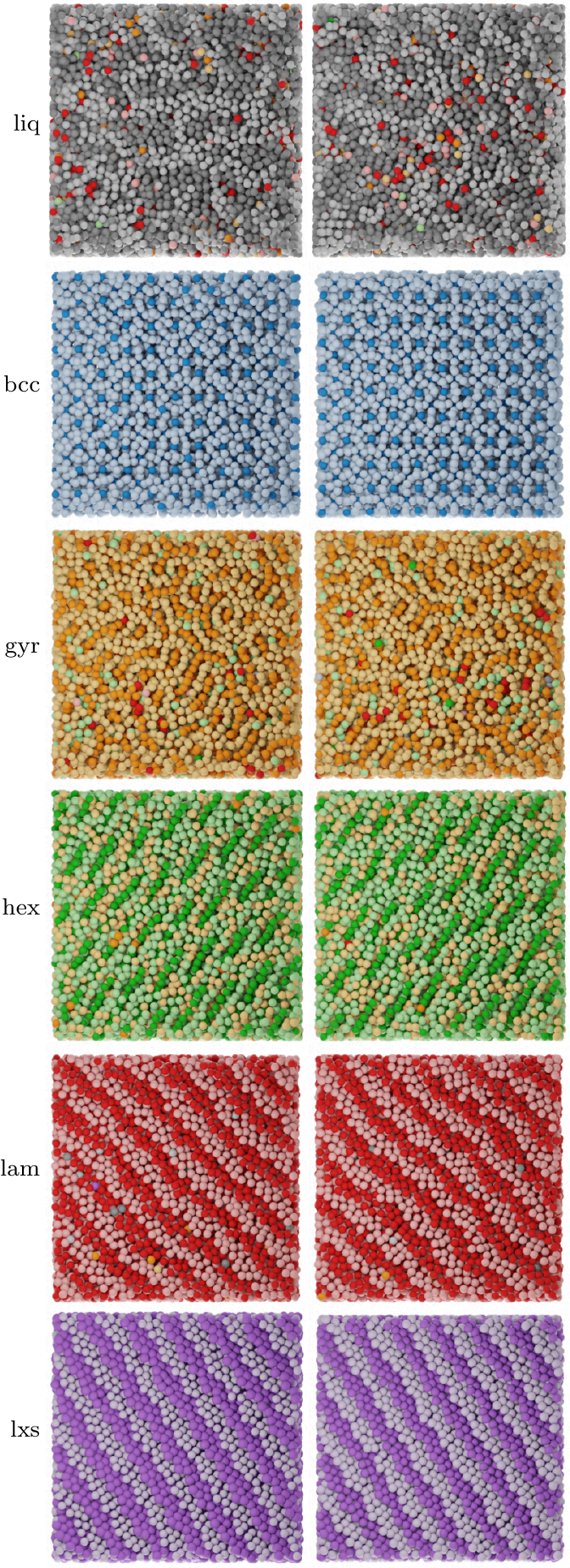}
\caption{
Top-down views of the mesophases as classified by the RFC model, using the categorical color scheme from Fig.~\ref{fig:meso-umap}.
Two replicas of each phase are shown to illustrate variability.
None of these environments were part of the RFC training set.
\label{fig:meso-snapshots}
}
\end{figure}

Finally, I show the results of classification by the RFC on entire snapshots (which are not included in the training data) in Fig.~\ref{fig:meso-snapshots}.
The bcc and lxs mesophases perform especially well due to their unusual structure compared to the others which shift their environments far away from the others in the UMAP.
The lamellar phase exhibits a few false negatives which are classified as liquid-like -- this is probably physical since the lamellar phase is well ordered in composition but not as much in space.
The liquid has a scattering of lamellar and gyroid false positives, which is consistent with values reported in the confusion matrices.
Finally, as already discussed, there is significant confusion over the gyroid and hexagonal phases.
While the bulk of the $B$-type particles are correctly classified, many of the hex-A particles are labeled as gyr-A, and some gyr-B are classified as lam-B.
It seems this method is quite reliable for classifying $B$-type particles, or large domains, but more uncertainty is introduced when classifying individual particles from these noisy mesophases.

\section{Conclusions}

I have introduced a new method for unsupervised learning of local atomic environments in molecular simulations based on simple rotation-invariant features and a familiar permutation-invariant pooling transformation.
These features are highly descriptive, generalize to multiple chemical species, and are human-interpretable.
The popular UMAP algorithm for unsupervised, nonlinear dimensionality reduction was applied to the resulting features to obtain a low-dimensional manifold for human interpretation (e.g., color coding).
Based on these embeddings, further unsupervised learning can be applied to obtain collective variables describing collections of particles such as an entire simulation domain.
Alternatively, the embedding can be post-processed by a supervised classification scheme such as the Random Forest Classifier to obtain robust classification on very few training samples.

I demonstrated the method on colloidal crystallization, ice crystals, and binary mesophases to show its broad applicability.
For the colloidal crystals, unsupervised learning alone provided quantitative measures which corresponded with physical intuition, and each dimension of the learned latent space yielded orthogonal insights into the details of the resulting microstructure.
With ice crystals, a supervised classifier was trained on a very small amount of data from the manifold (still learned by an unsupervised method) to directly compare against the supervised deep neural network approach used by Ref.~\cite{Defever2019} (PointNET).
Not only did the classification perform with comparable accuracy on substantially fewer training data, but my manifold learning scheme allowed the classifier to generalize crystal structures between bulk and surface-confined samples in a heterogeneous nucleation experiment, unlike PointNET (although this did not require new simulations, only additional data pre-processing).
A similar comparison was conducted for binary mesophases, with similar results -- my unsupervised manifold learning scheme provided a simplified topology for the classifier to learn on, reducing the volume of training data required to achieve a given accuracy.

While the particular features used here yield informative results, it is not clear from this preliminary study whether similar performance could be achieved with fewer features;
the current scheme was designed primarily for generality and ease of use rather than computational efficiency.
Ideally the features used as inputs to the manifold learning scheme would require as little pre-processing as possible (to accelerate the calculation of those features) and would be as sparse as possible (to minimize the cost of computing both the features and the embedding).
These considerations become especially important when the collective variables are used in advanced sampling schemes where they would need to be computed very frequently, or in optimization schemes requiring the use of gradients such as topology optimization.
As such, an ablation study on the features and the Gaussian expansion would be appropriate as future work.
In addition, a systematic study of the sensitivity to defects, particle deletions, and thermal fluctuations in solids near the melting point would help quantify whether using the full set of high-dimensional features provides benefits over pruned or hand-crafted feature vectors.
A similar strategy as the one used here should also be performed using some standard definitions of the atomic fingerprints as described in Section~\ref{sec:intro}.

The method presented here provides an incredibly versatile strategy to characterize and classify local atomic environments;
the combination of high-dimensional rotation- and permutation-invariant features with nonlinear manifold learning offers a flexible framework which in principle could describe nearly any local environment.
Unlike the graph-based measures I have employed for unsupervised manifold learning in the past \cite{Reinhart2017Machine, Reinhart2017Multi, Reinhart2018Automated}, these features are also differentiable, offering the possibility of efficient inversion from the latent space (i.e., backpropagation).
Notably, while this scheme gives favorable classification performance compared to deep neural networks, it can be deployed on even a handful of atomic environments without any labels or \textit{a priori} knowledge.
Thus, it may be employed in cases where the ground state structures are known or unknown, where the chemistry is uniform or heterogeneous, and where ordering occurs over one or many coordination shells.

\section*{Data availability} \label{sec:data}

The data required to reproduce these findings are available from the author upon reasonable request.

\section*{Acknowledgments}

I thank Sapna Sarupria and Steven Hall for providing the data used to train their PointNET model and for helpful discussions regarding the results.
I also thank Antonia Statt for helpful discussions and proofreading of the manuscript.
This work was supported in part by the Institute for Computational and Data Sciences and Materials Research Institute at the Pennsylvania State University.

\bibliography{refs}

\end{document}